\documentclass[modern]{aastex63}

\renewcommand{\deg}{^{\circ}}

\received{}
\revised{}
\accepted{}
\submitjournal{ApJ}

\shorttitle{3D Topology of an MFR}
\shortauthors{Hu et al.} 

\graphicspath{{FIGS_JUNE14_2012/}}
\begin{document}

\title{Validation and interpretation of three-dimensional configuration of a magnetic cloud flux rope}

\correspondingauthor{Qiang Hu}
\email{qiang.hu@uah.edu, qiang.hu.th@dartmouth.edu}


\author[0000-0002-7570-2301]{Qiang Hu}
\affiliation{Department of Space Science \\
Center for Space Plasma and Aeronomic Research (CSPAR) \\
The University of Alabama in Huntsville \\
Huntsville, AL 35805, USA}

\author[0000-0003-3218-5487]{Chunming Zhu}
\affiliation{Physics Department,
Montana State University,
Bozeman, MT 59717, USA}
\author[0000-0001-8749-1022]{Wen He}
\affiliation{Department of Space Science \\
The University of Alabama in Huntsville \\
Huntsville, AL 35805, USA}

\author{Jiong  Qiu}
\affiliation{Physics Department,
Montana State University,
Bozeman, MT 59717, USA}

\author[0000-0002-6849-5527]{Lan K. Jian}
\affiliation{NASA Goddard Space Flight Center, Greenbelt, MD 20771, USA}

\author[0000-0003-0819-464X]{Avijeet Prasad}
\affiliation{Institute of Theoretical Astrophysics, \\
University of Oslo, Postboks 1029 Blindern, 0315 Oslo, Norway}
\affiliation{Rosseland Centre for Solar Physics, \\
University of Oslo, Postboks 1029 Blindern, 0315 Oslo, Norway}

%
%
%
%
%
%
%



\begin{abstract}

 One  ``strong" magnetic cloud (MC) with the magnetic field magnitude reaching $\sim$ 40 nT at 1 au during 2012 June 16-17  is examined in association with  a pre-existing magnetic flux rope (MFR) identified  on the Sun.
The MC is characterized by a quasi-three dimensional (3D) flux rope model based on in situ measurements from the Wind spacecraft. The magnetic flux contents and other parameters are quantified. In addition, a correlative study with the corresponding measurements of the same structure crossed by the Venus Express (VEX) spacecraft at a heliocentric distance 0.7 au and with an angular separation $\sim 6^\circ$ in longitude is performed to validate the MC modeling results. The spatial variation between the Wind and VEX magnetic field measurements is attributed to the 3D configuration of the structure as featured by a knotted bundle of flux.
The comparison of the magnetic flux contents between the MC and the source region on the Sun indicates that the 3D reconnection process accompanying an M1.9 flare may correspond to the magnetic reconnection between the  field lines of the pre-existing MFR rooted in the opposite polarity footpoints. Such a process  reduces the amount of the axial  magnetic flux in the  erupted flux rope, by approximately 50\%, in this case. 

\end{abstract}



\section{Introduction} \label{sec:intro}

Coronal mass ejections (CMEs) are one important type of solar eruptions that are closely related to solar flares. They can have long-lasting impacts that may manifest throughout the interplanetary space. Both remote-sensing and in situ spacecraft observations are available during the initiation, eruption and propagation stages of a CME (and sometimes the accompanying flare). When a CME reaches  the interplanetary space, it is called an interplanetary CME (ICME) with a variety of distinctive signatures present in the in situ data \citep{2006SSRv..123...31Z}. The magnetic field structure embedded or hypothesized to have formed at early stages of a CME eruption is believed to be directly associated with the various manifestations of the corresponding flare/CME evolution and eruptions.

A magnetic flux rope (MFR) is generally believed to form the core structure of a CME eruption \citep[e.g.,][]{Vourlidas2014,2018Natur.554..211A,2020Liurui,2021Jiang}. However the existence or the formation of such a structure on the Sun has yet to be fully elucidated. A clear definition for an MFR, either intuitively or preferably based on magnetic field properties, has to be articulated and agreed upon. Admittedly, despite many indirect observational signatures of MFRs on the Sun \citep[e.g.,][]{2020Liurui}, it remains challenging for direct magnetic field measurements of an MFR, especially in the corona. On the other hand, such direct measurements are available in the interplanetary space, taken when an ICME encounters an observing spacecraft. In particular, one class of ICMEs, so-called magnetic clouds (MCs), possesses a unique set of signatures in the in situ magnetic field and plasma parameters \citep{1982GeoRL...9.1317B,1988JGR....93.7217B,1991burlaga}: 1) enhanced magnetic field magnitude, 2) smooth/gradual rotation in one or more components of the magnetic field, and 3) depressed proton temperature or $\beta$ value (the ratio between the plasma pressure and the magnetic pressure). Given these signatures and the fact that the spacecraft traverses the body of an MC structure, a flux rope configuration has been hypothesized to characterize the magnetic field structure of an MC since early times \citep[e.g.,][]{1990GMS....58..343G}. 

Based on in situ quantitative measurements and the flux rope hypothesis, various models are devised to fit the data. Among them, an early and commonly used one is based on a cylindrically symmetric linear force-free field (LFFF) configuration described by a simple analytic solution, so-called Lundquist solution \citep{lund}. The justification for an LFFF formulation is provided by the usually small $\beta$ ($\ll 1$) for an MC interval. The Lundquist solution represents a type of one-dimensional (1D) models which only have  spatial dependence on the radial distance from the cylindrical axis. Being the earliest quantitative approach, the Lundquist solution model has been widely applied to fit the magnetic field profiles from the in situ measurements of MCs \citep[e.g.,][]{1988JGR....93.7217B,Lepping1990,wucc}. An alternative and unique two-dimensional (2D) model was later  developed and applied to MC/ICME events, based on the Grad-Shafranov (GS) equation describing a 2D magnetohydrostatic equilibrium \citep{2001GeoRLHu,2002JGRAHu,Hu2017GSreview}. In the GS-based method, the force-free condition is no longer needed, and the solution is fully 2D over a cross section plane which does not change along the third (axial) dimension in the direction perpendicular to the plane. 

To complement and overcome the limitation of the 2D geometry of the GS reconstruction, we recently adopted a quasi-three dimensional (3D) model based on a more general LFFF formulation \citep{freidberg} which introduces much greater spatial variation than the Lundquist solution. The approach is to fit such an analytic model to in situ spacecraft measurements of the magnetic field components with uncertainty estimates (typically in the order of 1 nT on average for data at 1 au) through a formal least-squares  $\chi^2$ minimization algorithm \citep{Hu2021a,2021Husolphys}. The fitting results with a minimum reduced $\chi^2$ value around 1 are considered optimal, together with a set of geometrical and physical parameters characterizing a flux rope configuration with 3D features. It has been shown that the field lines exhibit typical twist along one dimension, as well as apparent writhe, in the form of the overall winding of a flux bundle corresponding to one major polarity. There are also cases with two bundles of mixed (opposite) polarities winding around each other \citep{Hu2021a}. Such a model complements the existing GS method and allows us to perform MC analysis with an additional tool, thus to expand our MC event databases.
In addition to the added features of a 3D configuration, the 3D model also tends to enable the selection of a larger MC interval  for analysis as compared with that for the 2D GS method to better reconstruct the underlying structure in its entirety. 

Since in situ MC modelings are nearly all based on single-point (or equivalently single-line) measurements across an MC structure, the validation of the model output is not always within reach. One way is to use more than one set of in situ spacecraft measurements obtained when the structure traverses multiple (often two) spacecraft with appropriate separation distances. Then the modeling result derived from one spacecraft dataset can be used to produce a ``prediction" or the expected values of the magnetic field along the path of the other spacecraft across the same structure. Such predicted values then can be compared with the actual measured ones to provide validation to the MC model. Such occasions are generally rare and such a validation approach has been carried out for the 2D GS reconstructions of a handful of MC events \citep{Hu2005,2008AnGeoM,2009SoPhM}. The latest applications to the quasi-3D model outputs were performed as well \citep{2021Husolphys,2021Hu2}. For example, in \citet{2021Hu2}, an MC structure was observed by both the Solar Orbiter (SO) at a heliocentric distance $\sim$0.8 au and Wind spacecraft at Earth with a longitudinal separation angle $\sim 4^\circ$. The analyses of the MC structure with both the 2D GS reconstruction and the quasi-3D model fitting were carried out by using the Wind spacecraft measurements and the expected values of magnetic field along the SO spacecraft path were produced for both models. The comparison with the actual measurements at SO yielded a linear correlation coefficient $>0.9$ for both methods. In \citet{2021Husolphys}, a similar validation study for the quasi-3D model between the Advanced Composition Explorer (ACE) and the Solar and TErrestrial RElations Observatory (STEREO) B spacecraft, separated by $\sim 3^\circ$ in longitude near 1 au, yielded a correlation coefficient 0.89. 

The other way of validation, still by employing multi-spacecraft measurements, is to relate the in situ MC flux rope properties with their counterparts in the corresponding solar source regions derived from multi-wavelength remote-sensing observations. Early attempts, through rigorous quantitative analyses of both in situ MCs by the 2D GS method and the corresponding source region properties in terms of the magnetic flux contents, were made by \citet{Qiu2007,2014ApJH} for about two dozen events with flare-CME-ICME/MC associations. It was found that the magnetic flux contents of MCs in terms of the toroidal (axial) and the poloidal flux correlate with the magnetic reconnection flux derived from the accompanying flare ribbon observations \citep{Qiu2004,Qiu2010}. In particular the comparison between  the poloidal flux and the reconnection flux exhibits a one-to-one correspondence \citep{2014ApJH,2017Gopalswamy}. Additional case studies also followed to further relate the axial flux contents in MCs to their source regions, either with or without the direct identification of the corresponding MFR footpoints on the Sun \citep{2017NatCo...8.1330W,Xing_2020}.  These  results support the scenarios envisaged by \citet{Longcope2007a} and \citet{2017SoPh..292...25P} \citep[see, also, ][]{1989vanBallegooijen} of MFR formation via sequential magnetic reconnection between sheared magnetic loops, often manifested as ``two-ribbon" flares. During the process, the reconnection flux is largely injected into the ensuing MFR forming above the flare loops and ejected with the CME eruption. The brightened flare ribbons map the footpoints of reconnected field lines. Combined with the corresponding magnetograms, the amount of magnetic flux encompassed by the ribbon areas (equivalent to the reconnection flux) can be measured routinely through standardized procedures \citep[e.g.,][]{2017Kazachenko}. Such quantitative intercomparison not only provides insight into the MFR formation on the Sun, but also results in (indirect) validation to the in situ MC modeling results, given implied consistency with certain theoretical frameworks. In the above mentioned works, the scenario of an MFR formation through magnetic reconnection is supported by the analysis result that the amount of reconnection flux corresponds well to the poloidal flux of the MFR with an approximate 2D geometry. 

Needless to say, the complexity in the modeling of solar source region magnetic field topology has outpaced the in situ modeling of MCs. We attempt to develop and apply a more complex model that is on par in complexity with the source region magnetic field topology, and to explore connections with more non-standard processes that go beyond a standard view, e.g., under an approximate 2D geometry, for flare associated reconnection \citep[e.g.,][]{2019Aulanier}. Therefore the motivation of the current study is two-fold: 1) to validate the quasi-3D MC model by employing aforementioned validation approaches, and 2) to relate to the source region properties involving 3D reconnection processes. Therefore we aim to study another view on flare-CME-ICME connection based on an alternative interpretation of one event analysis result.

We select the solar event on 2012 June 14 (SOL2012-06-14) for the present study. The sequence of events includes a series of confined C-class flares preceding the main M1.9 flare, followed by the accompanying CME eruption. The corresponding remote-sensing observations are obtained from the Solar Dynamics Observatory (SDO) including Atmospheric Imaging Assembly (AIA) and Helioseismic and Magnetic Imager (HMI), and the  coronagraphs onboard Solar and Heliospheric Observatory (SOHO) and STEREO spacecraft. They  had been analyzed by \citet{2019ApJ...871...25W} and \citet{zhu2020} in great details. Prior to the eruptions, an MFR structure was inferred from conjugate dimming signatures and both footpoint regions were clearly identified and found to root in strong and opposite magnetic polarity regions with strong vertical currents  \citep{2019ApJ...871...25W}. This pre-existing MFR prior to the main flare and CME eruptions was found to be formed via a sequence of reconnections facilitated by photospheric and coronal evolution processes \citep[see, also][]{2017SoPh..292...71J}. Its magnetic properties were derived by \citet{2019ApJ...871...25W} and will be compared with our in situ modeling results. The erupted CME reached Earth on June 16 and lasted for about a day as observed by the Wind spacecraft. Around that time period, the Venus Express (VEX) spacecraft at Venus was in approximate radial alignment with Earth and detected the same ICME/MC structure, thus providing an additional set of in situ measurements (magnetic field only; \citet{ZHANG20061336}) for the validation study. In what follows, we will present the unique analysis of the MC structure by the quasi-3D model based on the Wind spacecraft measurements, complemented with a correlative study with the corresponding VEX measurements. Based on these new results and their connection to the source region properties, mainly associated with the pre-existing MFR and the measured reconnection flux, we offer an interpretation of the evolution of the MFR topology upon eruption, on the basis of a quantitative analysis of the derived toroidal (or axial) magnetic flux, which is still well characterized under 3D geometries. And increasingly it has been well quantified from solar observations \citep[e.g.,][]{2020Innov...100059X}.

This paper is organized as follows. In Section~\ref{sec:event}, we present an overview of the event including the timelines for the associated flare, CME and ICME, based on prior works and focusing on in situ measurements and identification of an MC interval. We present the new modeling results for the MC by applying the quasi-3D  model to the in situ Wind spacecraft data. In particular, a correlative study with the corresponding VEX in situ measurements is performed. We relate the in situ MC modeling results with the solar source region properties in Section~\ref{sec:interpretation} and offer an interpretation of such a quantitative connection. Finally we conclude in the last section.

\section{Event Overview and In situ MC Model Result}\label{sec:event}
\begin{figure}
\centering
\includegraphics[width=8.3cm]{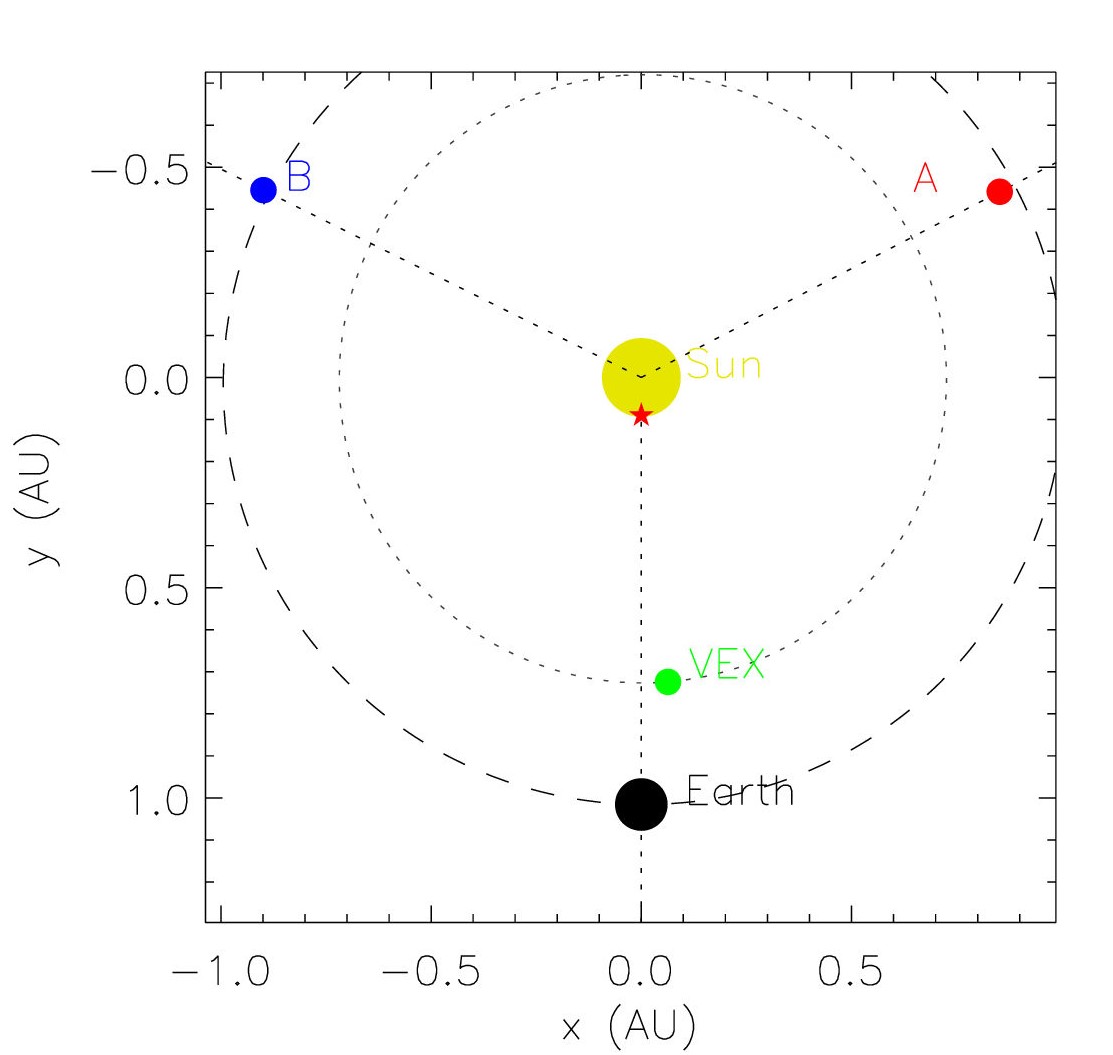}
\caption{The spacecraft locations projected on the ecliptic plane on  2012 June 14: STEREO A and B, VEX, and Wind (Earth), as denoted. The red star indicates the approximate location of the solar source region of the flare/CME eruption (Earth facing in this case). }\label{fig:sc}
\end{figure}

\subsection{Event Overview and In Situ Measurements}\label{subsec:insitu}
The flare-CME event on 2012  June 14 has been well studied and the observational analyses of remote-sensing measurements were conducted and reported in several prior studies \citep[e.g.,][]{2019ApJ...871...25W}. In particular, \citet[][and references therein]{2017SoPh..292...71J} have examined the association among the flares, CME, and the corresponding ICME thoroughly with a set of comprehensive remote-sensing observations including UV/EUV, coronagraphic, and microwave imageries. We refer readers to those references for details and only provide a brief summary of the event sequence, but focus on in situ measurements and  MC modeling in this section. The pair of an M1.9 flare (peaking around 14:30 UT in the soft X-ray flux) and a halo CME (appearing around 14:12 UT in SOHO/LASCO C2) eruptions occurred on  2012 June 14, followed by an ICME/MC passage at 1 au during  June 16-17 \citep{2017SoPh..292...71J}. In addition to multiview remote-sensing observations from SDO, SOHO, STEREO A and B spacecraft, the ICME also passed the Venus Express (VEX) spacecraft near Venus at a heliocentric distance $r_h\approx 0.7$ au, when VEX was nearly radially aligned with the Sun-Earth line \citep{Chi_2020,Kubicka_2016}.

\begin{figure}
\centering
\includegraphics[width=8.3cm]{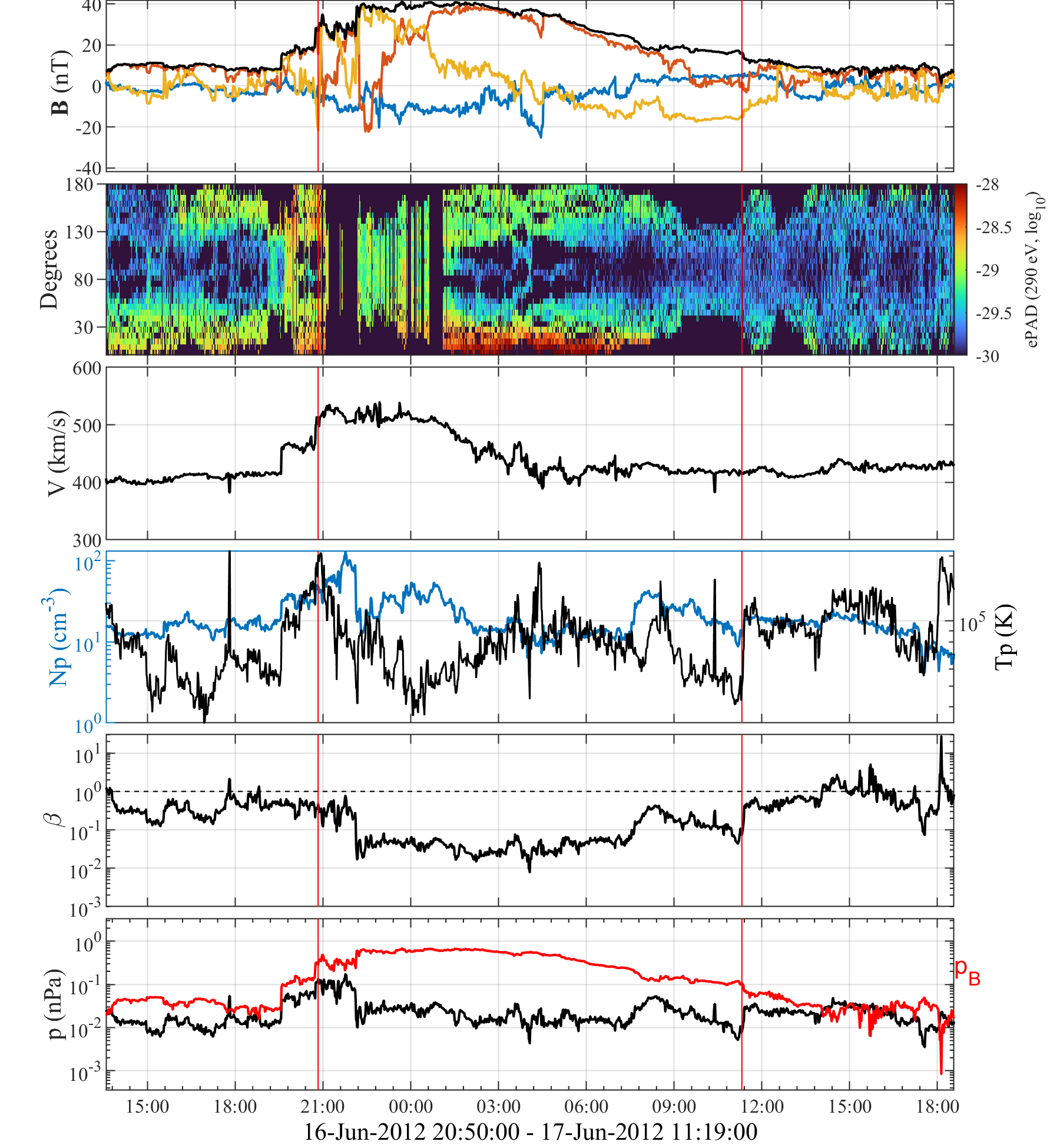}
\caption{In situ time-series measurements of magnetic field and plasma parameters from the Wind spacecraft. From the top to  bottom panels are the magnetic field components in the GSE-X (blue), Y (red), and Z (gold) coordinates, and magnitude (black), the ePAD at 290 eV energy channel, the solar wind speed, the proton density and temperature, the proton $\beta$, and the corresponding  proton pressure (black) and the magnetic pressure (red). The vertical lines mark the MC interval given beneath the bottom panel for analysis.  }\label{fig:dataplot}
\end{figure}

\begin{figure}
\centering
\includegraphics[width=8.3cm]{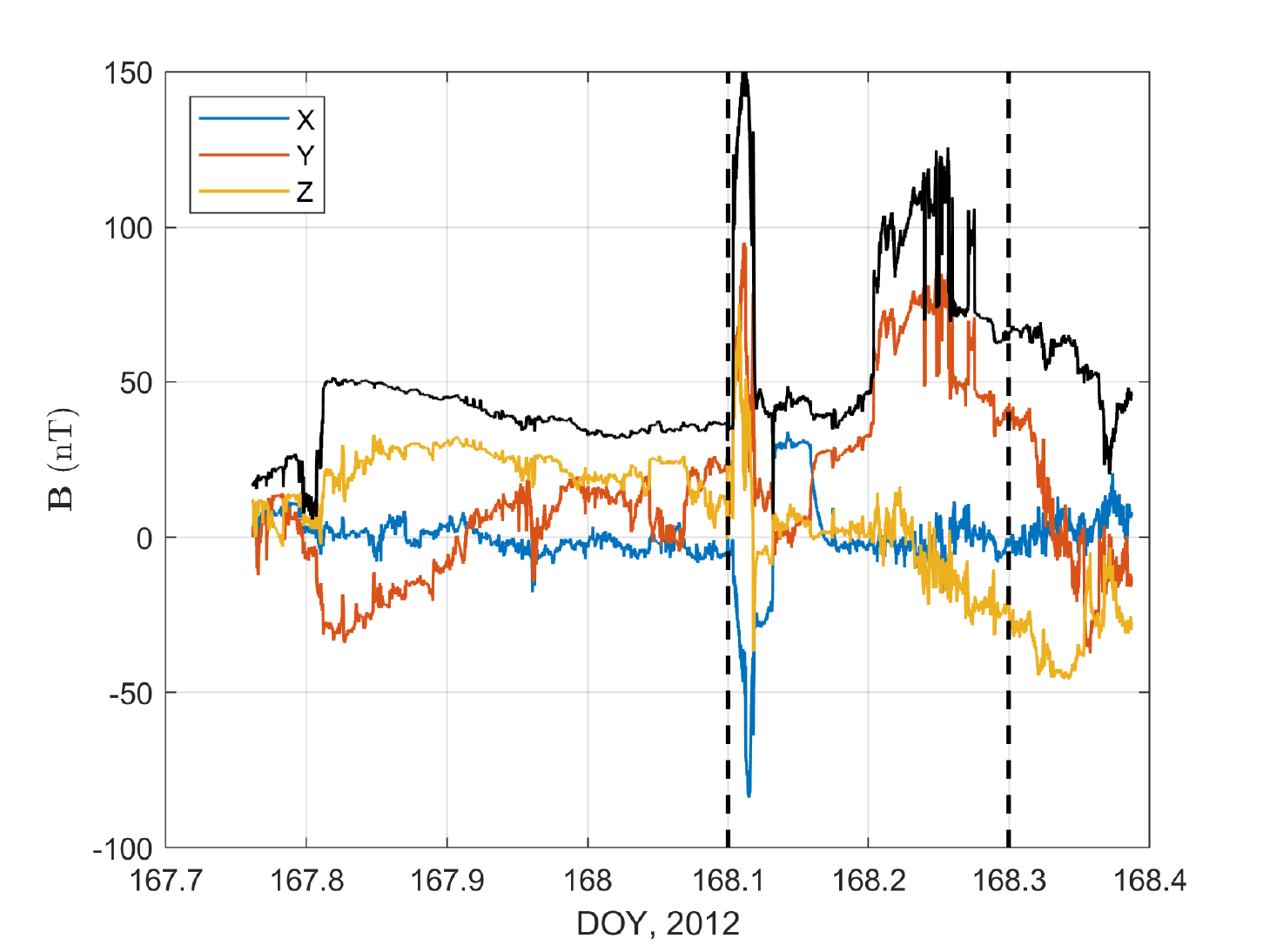}
\caption{The time series of magnetic field components and magnitude (black) in the Venus Solar Orbital (VSO) coordinates 
(the equivalence of the GSE coordinate system defined for Venus) for the interval of the MC passage at Venus during  2012 June 15 - 16, approximately between day of year (DOY) 167.7 and 168.4. The dashed lines mark the interval  approximately between DOY 168.1 and 168.3 when   VEX exited  into a different  regime (likely the Venusian magnetosphere) from the solar wind. See text for details. }\label{fig:BVex}
\end{figure}

Figure~\ref{fig:sc} shows the relative locations of multiple spacecraft  on the ecliptic  plane for this event. Both STEREO A and B were in approximate quadrature with respect to Earth, which provided side views toward the eruption of the corresponding CME and allowed for detailed analysis of the CME kinematics \citep{zhu2020}. The spacecraft near Earth including ACE and Wind detected the subsequent ICME/MC, while VEX, located at $r_h\approx 0.7$ au and about 6$^\circ$ away from the Sun-Earth line to the west, also provided the  magnetic field measurements of the same ICME structure about 1 day earlier.  
Figures~\ref{fig:dataplot}  and \ref{fig:BVex} show the corresponding in situ time-series measurements at the Wind and VEX spacecraft, respectively. At Wind, both the magnetic field  $\mathbf{B}$ and solar wind plasma parameters including flow velocity $\mathbf{V}$ (all in the GSE, Geocentric Solar Ecliptic, coordinate) are used for our analysis. In Figure~\ref{fig:dataplot}, several selected parameters are shown, including the suprathermal electron pitch angle distribution (ePAD) for the 290 eV energy channel. The magnetic field magnitude reaches nearly 40 nT at maximum, and the field components exhibit signatures of rotation within the marked interval. The solar wind speed shows a gradual decrease from $\sim$ 500 km/s to $\sim$ 400 km/s. The effect of such a change will be examined in Section~\ref{sec:FS}. The proton density and temperature vary with periods of depressed proton temperature within the interval. The resulting proton pressure displays little overall variation, but is significantly lower than the corresponding magnetic pressure, which results in significantly depressed proton $\beta$, with an average value 0.10 within the marked interval. Therefore we identify the marked interval as the passage of an MC and select the corresponding data segments with 1 minute resolution for the subsequent analysis. At VEX, only the magnetic field measurements were available and occasionally contaminated by the Venusian magnetic field, as indicated by large excursions of field magnitudes exceeding, e.g., 80 nT,  due to possible bow shock crossings \citep{Xu_2019}. Based on timing analysis (see Section~\ref{sec:FS-VEX}), the passage of the ICME most likely begins around DOY 167.8 with the field magnitude reaching about 50 nT, larger than the maximum magnitude at 1 au. The dashed lines mark the interval bounded by possible bow shock crossings, as indicated by the magnetic field of large magnitude and the associated  frequent and abrupt changes  unlike solar wind behavior.
In Section~\ref{sec:FS-VEX}, this marked time interval is therefore excluded from the correlative comparison with the MC model result derived from the Wind spacecraft measurements. 

It is worth noting that from an ICME catalogue compiled by \citet{chi2016statistical} for years 1995-2015 based on Wind and ACE spacecraft in situ measurements, the distribution of the average magnetic field intensity within identified ICMEs spans a range between a few nT to more than 37 nT, peaking around the mean value $\sim$10 nT. An average magnetic field magnitude greater than 25 nT is lying in the tail portion of the distribution with only a handful of such events ($<10$ out of a total number of nearly 500 events), indicating rare occurrences. This MC event has an average magnetic field magnitude of 29 nT and  falls within that range with strong magnetic field.
It was noted by \citet{shenfphy.2021.762488} that this event is among ``the top four strongest ICME [sic] in magnetic field strength, according to their ICME catalogue\footnote{\url{http://space.ustc.edu.cn/dreams/wind_icmes/}; See also \citet{chi2016statistical}.}".  
This fact is consistent with our identified connection of this MC flux rope structure with the  MFR rooted in strong magnetic field regions with opposite polarities on the Sun prior to the main flare (M1.9) eruption, as elucidated by \citet{2019ApJ...871...25W}. It was found that the average vertical magnetic field strength for the identified positive and negative polarity footpoint regions was 1555$\pm35$ G and -710$\pm45$ G, respectively. We will further interpret such a connection in Section~\ref{sec:interpretation}.

\begin{table}[tbh]

    \caption{Summary of geometrical and physical  parameters for the MC from the FS model based on Wind spacecraft measurements.}\label{tbl:case1}
\centering
    \begin{tabular}{ccccccc}
        \hline
         $B_{z0}$ (nT) & $C$& $k$ & $\alpha$ & $\hat{\mathbf{z}}=(\delta, \phi)$\tablenotemark{a} &  $\Phi_z$ (Mx)& Chirality\\
\hline
    38 &-0.7138& -1.286 &2.626 & (56, 150)   & 8.0-14& $\mathbf{+}$\\
             &$\pm 0.3518$&$\pm$0.5123 &$\pm$0.2766&$\pm (10,15)$  & $\times10^{20}$ &(right-handed) \\
    \hline
    \end{tabular}\\ \tablenotetext{a}{The polar angle $\delta$ from the ecliptic north, and the azimuthal angle $\phi$ measured from GSE-X towards GSE-Y axes, both in degrees.}
\end{table}

\subsection{A Quasi-3D MC Model: Freidberg Solution}\label{sec:FS}
For the analysis of the MC interval by employing the Wind spacecraft in situ  data, we apply a newly developed approach \citep{Hu2021a}, the optimal fitting to the Freidberg solution (FS), which describes a quasi-3D magnetic field configuration. As given below, the three magnetic field components of an FS model in  cylindrical coordinates $(r,\theta,z)$ satisfy the LFFF formulation, $\nabla\times\mathbf{B}=\alpha\mathbf{B}$ $(\alpha\equiv Const)$, but exhibit variations in all three spatial dimensions \citep{freidberg},
\begin{eqnarray}
\frac{B_z(\mathbf{r})}{B_{z0}} & = & J_0(\alpha r)+CJ_1(l r)\cos(\theta+kz), \label{eq:B}\\
\frac{B_\theta(\mathbf{r})}{B_{z0}} & = & J_1(\alpha r)-\frac{C}{l}\left[\alpha J'_1(l r)+\frac{k}{l r}J_1(l r)\right]\cos(\theta+kz), \\
\frac{B_r(\mathbf{r})}{B_{z0}} & = & -\frac{C}{l}\left[k J'_1(l r)+\frac{\alpha}{l r}J_1(l r)\right]\sin(\theta+kz). \label{eq:B4}
\end{eqnarray}
Here the constant force-free parameter is denoted $\alpha$. Additional constant parameters are $C$, $k$, and $l=\sqrt{\alpha^2-k^2}$. The normalization constant $B_{z0}$ is pre-determined and taken as the maximum magnitude among all three magnetic field components within the MC interval. The functions, $J_0$ and $J_1$, are the usual Bessel functions of the first kind of order 0 and 1, respectively. It is clear that the amplitudes of variations in $\theta$ and $z$ (both periodic but adding 3D features) are controlled by the parameter $C$. For $C\equiv 0$, it reduces to the commonly known Lundquist solution \citep{lund} with only $r$ dependence. 

An optimal fitting between the measured magnetic field components and those yielded by the analytic solution represented by equations~(\ref{eq:B})-(\ref{eq:B4}) along the spacecraft path across the MC structure is performed to derive the set of free parameters. They include $C$, $k$, and $\alpha$, and additionally the orientation of the local cylindrical axis $z$. An algorithm, following the least-squares approach described by \citet{2002nrca.book.....P} including measurement uncertainty estimates, has been implemented and applied to a few event studies \citep{Hu2021a,2021Husolphys}. It has succeeded in yielding a minimum reduced $\chi^2$ value around 1, and the associated ``goodness-of-fit" metric $Q>10^{-3}$, to be considered acceptable \citep{2002nrca.book.....P}. 

For this MC interval, the average proton $\beta$ is about 0.10. A reference frame  in which the structure appears to be stationary is determined as the deHoffman-Teller (HT) frame with a constant velocity, $\mathbf{V}_{HT}=[-467.12, -9.41, -20.76]$ km/s in the GSE coordinate system based on Wind spacecraft in situ measurements (see \citet{2021Husolphys} for details on the justification for the use of $\mathbf{V}_{HT}$ and  the associated HT analysis). In short, in such a frame, the spacecraft is moving across the structure with the velocity $-\mathbf{V}_{HT}$. The ratio between the remaining flow $\mathbf{v}'=$ $\mathbf{V}-\mathbf{V}_{HT}$ and the local Alfv\'en velocity is evaluated by the slope (so-called Wal\'en slope) of the regression line between the components of the two velocities, as an indication of the relative importance of the inertial force compared with the Lorentz force. For this event interval, the Wal\'en slope is -0.13 with the magnitude much less than 1. Therefore the inertial force can be considered negligible and a force-free equilibrium is considered valid. 

Table~\ref{tbl:case1} summarizes the main fitting parameters and derived quantities, where the parameters $k$ and $\alpha$ become dimensionless  by multiplying a normalization length $a$ \citep{2021Husolphys}. The orientation of the $z$ axis is expressed in terms of the two directional angles, and the axial magnetic flux $\Phi_z$ is obtained over the cross section plane perpendicular to $\hat\mathbf{z}$ within an area with $B_z>0$ \citep{2021Husolphys}. For $\alpha>0$, it has a positive sign of helicity (or right-handed chirality). The dimensionless parameter $\alpha$ is related to a twist number $\tau_0=\alpha/2$, the number of twist for a field line of length $2\pi a$, following \citet{2016LiuR}, for such an LFFF configuration with a constant $\alpha$. Namely, the twist number for a field line of length $L$ is calculated  by the following line integral along each individual field line,
\begin{equation}
    T_w=\int_L\frac{(\nabla\times\mathbf{B})\cdot\mathbf{B}}{4\pi B^2}dl=\frac{\alpha L}{4\pi}\propto L \quad (\alpha\equiv Const). \label{eq:Tw}
\end{equation}
By applying a length normalization, i.e., replacing $\alpha a$ by $\alpha$, it is obtained $\tau_0=\alpha/2$ for $L=2\pi a$.
It can be scaled for any field lines of arbitrary lengths in the configuration represented by an FS model.

\begin{figure}
\centering
\includegraphics[width=8.3cm]{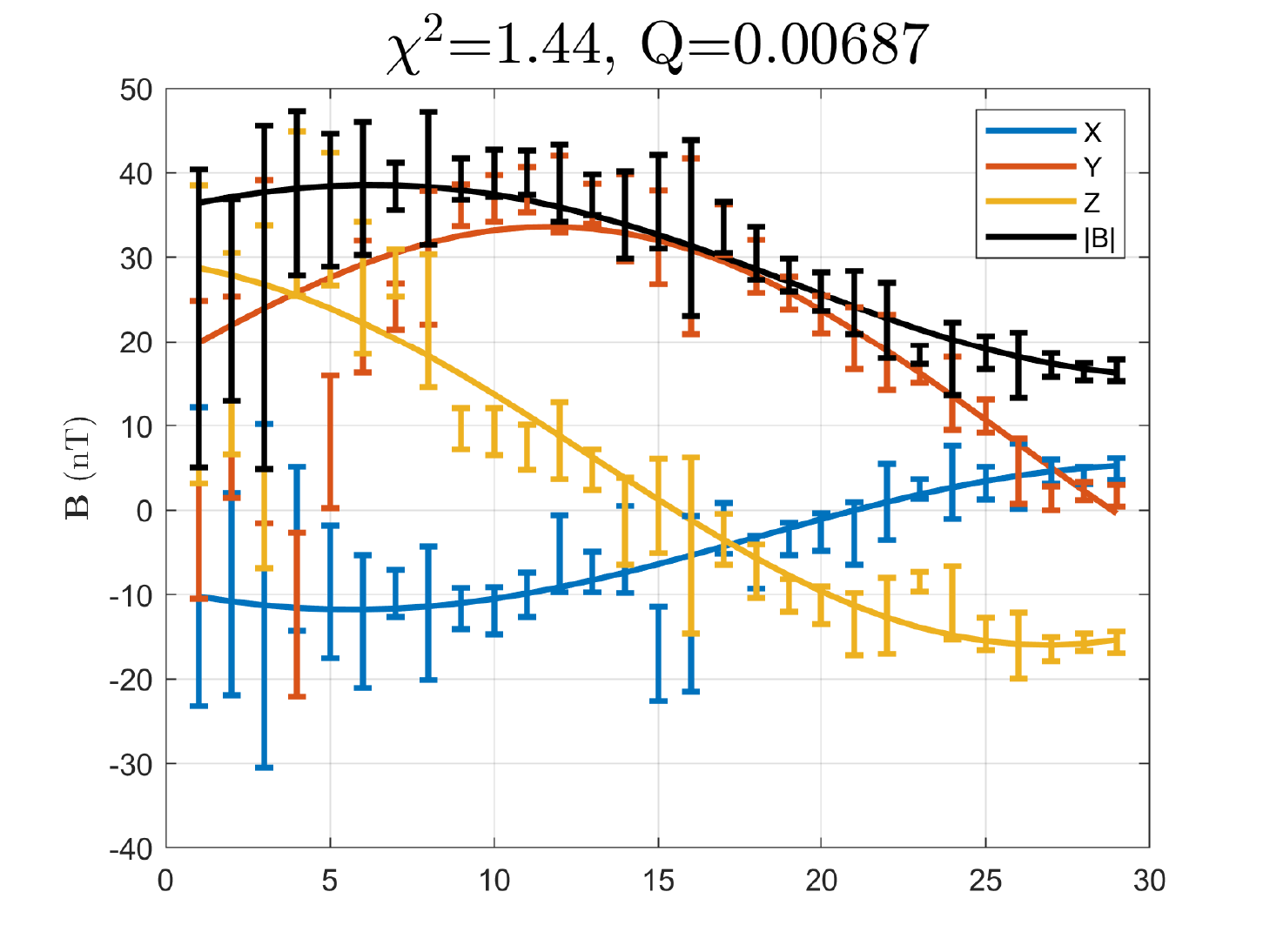}
\caption{The optimal fitting result for the MC interval by the FS model along the Wind spacecraft path (horizontal axis shows the indices of the data points). The magnetic field components and magnitude are shown for both the Wind spacecraft measurements with uncertainties as discrete errorbars and the corresponding FS model output by the solid curves (see the legend), both in the GSE coordinates. The minimum reduced $\chi^2$ value and the goodness-of-fit parameter $Q$ are indicated on top. }\label{fig:Brtn}
\end{figure}

Figure~\ref{fig:Brtn} shows formally the optimal fitting result to the FS formulation given by equations~(\ref{eq:B})-(\ref{eq:B4}) for the magnetic field measurements downsampled to about 30-minute resolution with uncertainties along the Wind spacecraft path. 
The minimum reduced $\chi^2$ value is 1.44, and the corresponding $Q=0.00687$. The set of main parameters as presented in Table~\ref{tbl:case1} represents the optimal output from the fitting procedure with uncertainty estimates based on  90\% confidence limits \citep{2002nrca.book.....P,2021Hu2}. The parameter $C$ has a magnitude close to 1, indicating significant deviation from a cylindrically symmetric configuration, as we will demonstrate below. The orientation of the $z$ axis in directional angles $(\delta,\phi)$ is obtained in the same procedure with uncertainties. In turn, the axial magnetic flux $\Phi_z$ is estimated to be 8.0-14$\times10^{20}$ Mx, mainly subject to the uncertainty in the $z$ axis orientation.

\begin{figure}
\centering
\includegraphics[width=8.3cm]{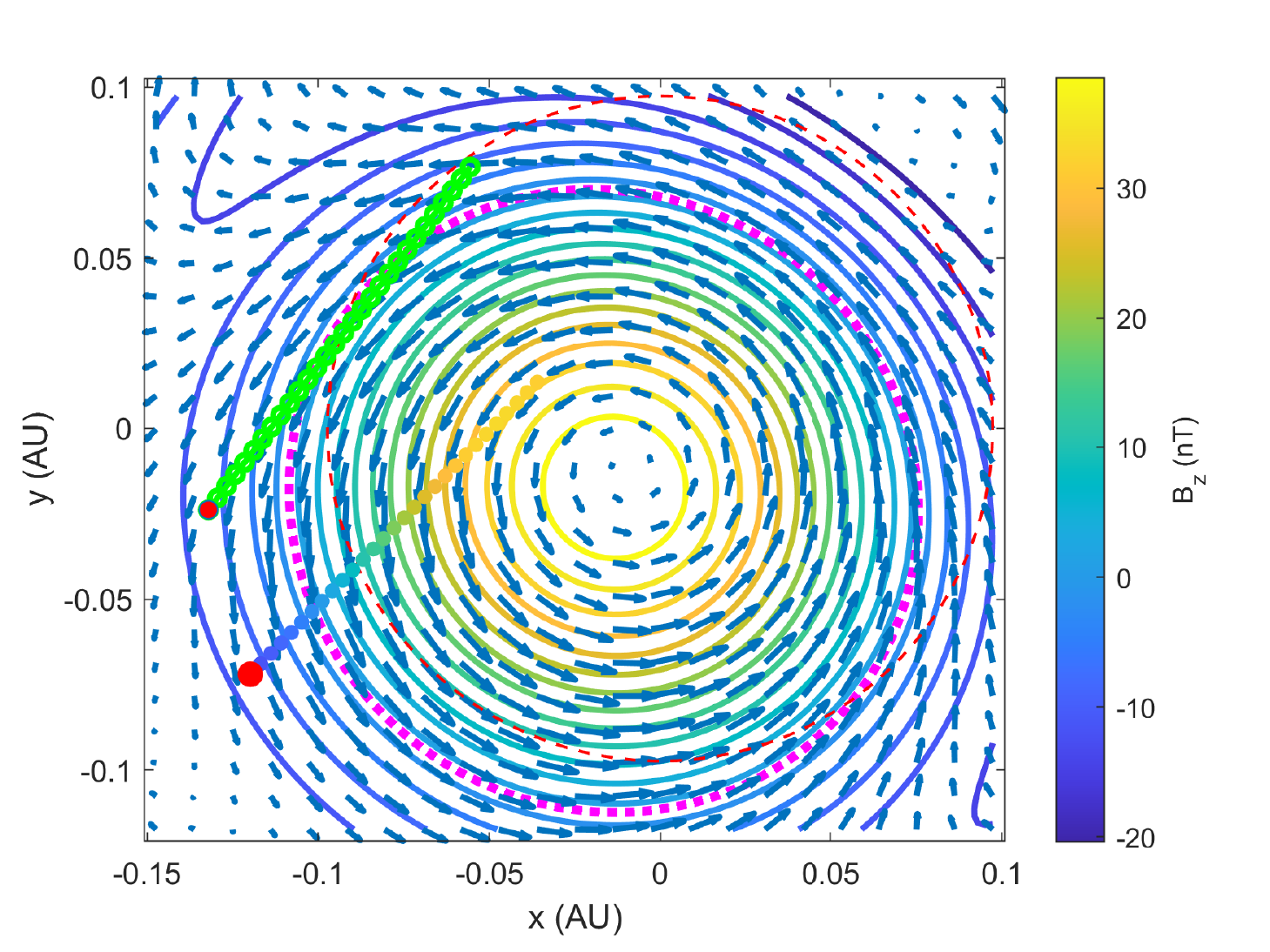}
\caption{The magnetic field configuration on one cross section of the FS model with the optimal fitting parameters. The blue arrows represent the transverse field components, and the axial field component $B_z$ is given by the color contours with color scales indicated by the colorbar. The thick dotted magenta curve marks the boundary where $B_z=0$. The thin red dashed circle is centered at the origin with the radius $a =  0.097$ au. The Wind and VEX spacecraft paths are projected separately by two groups of colored dots along two straight lines, with the former colored by the corresponding $B_z$ values and the latter in green. The end point along each path is marked in red color. }\label{fig:2Dview}
\end{figure}
Figure~\ref{fig:2Dview} shows a cross section of the fitted FS model at (arbitrarily selected) $z=0$. A boundary (closed in this view) is chosen where $B_z=0$ as illustrated by the thick dotted magenta curve, within which the axial field $B_z$ is positive, as indicated by the colorbar. This solution has a single dominant and positive $B_z$ polarity. Note that for this solution, unlike the 2D GS model, the transverse field as represented by the blue arrows is  no longer tangential to the contours of $B_z$, and such a cross section map changes with $z$. Both the Wind and VEX spacecraft paths are projected onto this view, although neither of the paths lies entirely on this plane. A correlative analysis between the FS model prediction along the VEX spacecraft path and the actual measurements of the magnetic field will be presented in Section~\ref{sec:FS-VEX}.

\begin{figure}
\centering
\includegraphics[width=.8\textwidth]{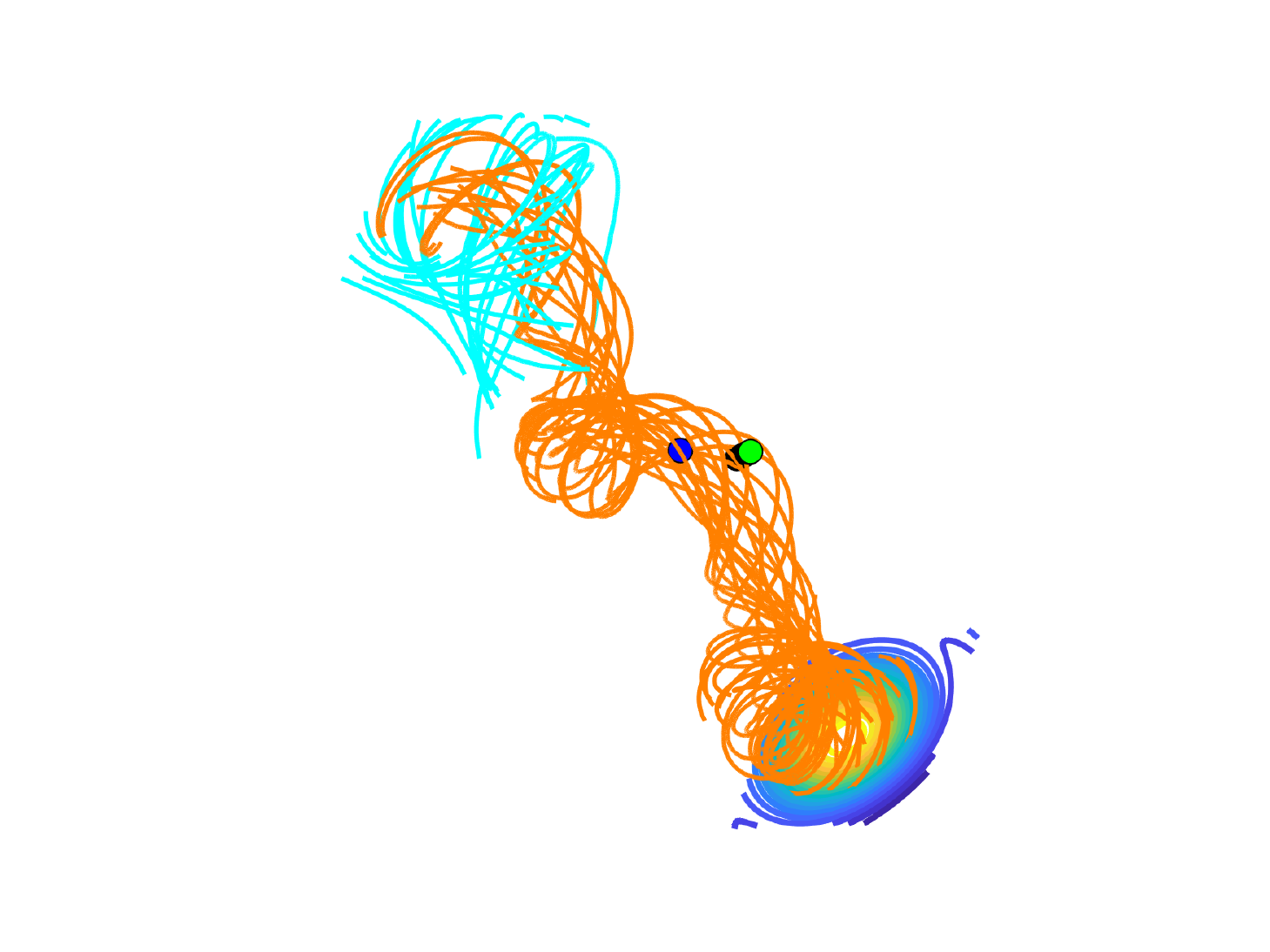}
\caption{The field-line plot in a Cartesian volume and in a view toward the Sun along the Wind spacecraft path. The ecliptic north is straight upward, and the east-west direction is horizontal. The orange lines are the field lines originating from the bottom plane where contours of $B_z$ are shown. The cyan lines originate from the top plane (not shown) and are going downward as well as out of the volume. Blue and green dots are the Wind and VEX spacecraft paths in this view, respectively. }\label{fig:3D}
\end{figure}
\begin{figure}
\centering
\includegraphics[width=.88\textwidth]{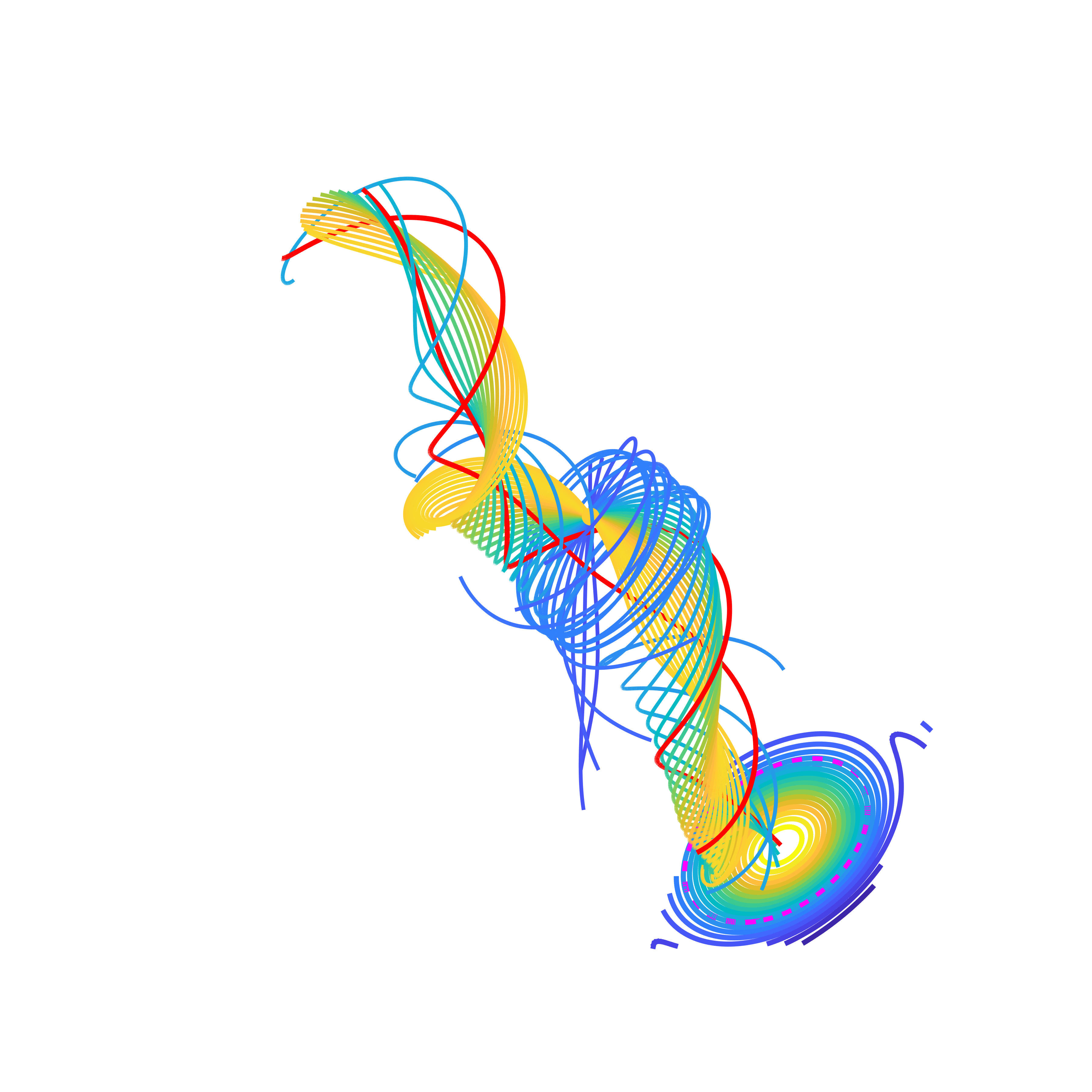}
\caption{Selected field lines crossing the Wind spacecraft path in the same  volume and view angle as Figure~\ref{fig:3D} and color coded by the corresponding $B_z$ values with the same colorbar as Figure~\ref{fig:2Dview}. Two thicker lines in red highlight the 3D nature of the configuration: one is rooted on the bottom plane where the contours of $B_z$ are shown, and starts from the point with the maximum $B_z$; the other crosses the point of maximum $B_z$ on a different plane as shown in Figure~\ref{fig:2Dview}. They apparently are not straight lines. 
}\label{fig:orb}
\end{figure}

The 3D nature of the FS model result is better illustrated by Figure~\ref{fig:3D}, where a 3D view of the field-line configuration from the Earth's perspective toward the Sun is given. The main flux bundle in orange color is rooted on the bottom cross-section plane with a major positive $B_z$ polarity and is winding upward along the positive $z$ axis direction. It displays a writhe or a knot in the body of the flux bundle, which gives rise to the 3D feature of the magnetic field configuration. In the same view, Figure~\ref{fig:orb} presents the selected field lines intercepting the Wind spacecraft path and color-coded by the corresponding $B_z$ field components with the same scales indicated by the colorbar in Figure~\ref{fig:2Dview}, except for the two red lines. Along the Wind spacecraft path, from the beginning to the end, the $B_z$ component changes from positive to negative values. This indicates the corresponding changes in the field line directions from going upward to downward, with respect to the bottom plane, as the color of the field lines transitions from bright gold to blue. The main flux bundle or band with gold to light blue colors twists collectively along the $z$ dimension. Overall there lacks a central straight field line in this configuration. This is further illustrated by highlighting the two field lines in red which connect to two ``central" locations with the maximum $B_z$ values at two different cross section planes (i.e., with two different $z$ values). Overall the field-line configuration showcases a ``twisted-ribbon" type of topology with a knotted appearance.

\subsection{Correlation with VEX Measurements}\label{sec:FS-VEX}
\begin{figure}
\centering
\includegraphics[width=\textwidth]{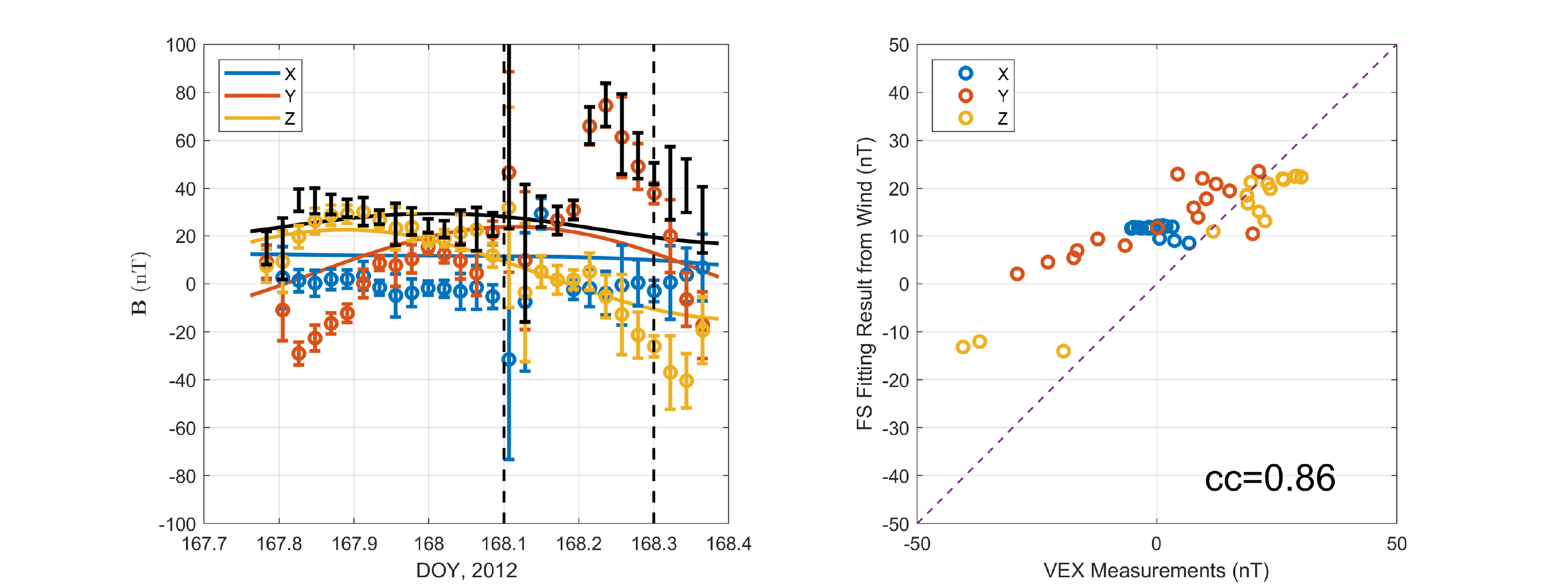}
\caption{Left panel:  time series of the magnetic field components (see legend) and magnitude (black) at VEX, including the VEX measurements represented by errorbars and the FS model output by smooth curves. The pair of dashed lines marks the period of close Venus encounter which is excluded from the analysis result shown in the right panel.  Right panel: the corresponding  scatter plot of the two sets of field components, yielding a correlation coefficient $cc=0.86$. The dashed line indicates the one-to-one diagonal line. }\label{fig:vexcorr}
\end{figure}

Both Wind and VEX spacecraft paths cross the main flux bundle, as indicated in Figures~\ref{fig:2Dview} and \ref{fig:3D}, by taking into account a nominal time shift between the two spacecraft, due to the relatively small separations in both heliocentric distances and longitudes. The time shift is calculated by considering a constant propagating speed of the structure at $|\mathbf{V}_{HT}|$. By omitting the temporal change of the structure during the propagation, we compare the magnetic field components along the VEX spacecraft path derived from the FS model based on the Wind spacecraft measurements with the actual VEX data shown in Figure~\ref{fig:BVex}. The comparison including the magnetic field magnitude is given in Figure~\ref{fig:vexcorr} with the VEX data downsampled to about 30-minute resolution with uncertainty estimates. The component-wise correlation is displayed in the right panel of Figure~\ref{fig:vexcorr}, yielding a linear correlation coefficient $cc=0.86$, for the two data sets excluding the data points between the vertical dashed  lines, many reaching large magnitudes beyond $ 40$ nT, in the left panel. Those are likely due to the contamination from the Venusian magnetosphere  as  discussed in Section~\ref{subsec:insitu}.

\begin{table}[h]
    \caption{Summary of results from both remote-sensing and in-situ data analyses of the source region and the corresponding MC properties.}\label{tbl:results}
    \centering
    \begin{tabular}{ccc}
        \hline
        Parameters & Source Region\tablenotemark{a}  & In-situ MC Model  \\
       (fluxes in $10^{20}$ Mx)            &      &                3D (FS)\\
\hline
        Axial magnetic flux $\Phi_z$  & (+)42, (-)30   & 8.0-14\\
        Total twist number $\tau$ & $\sim 2$ &  1.9-2.4 /au\\
        Magnetic helicity ($10^{42}$ Mx$^2$)  & 6-10 &  1.6-3.5 /au\\
        Reconnection flux & (+)$22\pm4.7$, (-)$21\pm5.3$ &  ...\\
    \hline
    \end{tabular}
    \tablenotetext{a}{The first three rows are estimates from \citet{2019ApJ...871...25W} for the pre-existing MFR, including the axial magnetic fluxes for the positive (+) and negative (-) polarity footpoint regions based on  dimming and magnetograph measurements.}
\end{table}


 \section{Connection to Solar Source  Properties and Interpretation}\label{sec:interpretation}

As we have discussed that based on prior studies \citep[e.g.,][]{2017SoPh..292...71J} and the fact that the MC as detected at 1 au possesses a significant amount of magnetic flux with unusually large magnetic field magnitude, the correspondence of the MC analyzed in Section~\ref{sec:event} to the pre-existing MFR identified by \citet{2019ApJ...871...25W} is one likely scenario. We base our comparison of a handful of characteristic physical properties between the MC and the solar source region on this connection. Again we intend to illuminate an alternative and perhaps unusual scenario that goes beyond the standard view as we described in Section~\ref{sec:intro}.

\begin{figure}
\centering
\includegraphics[width=8.3cm]{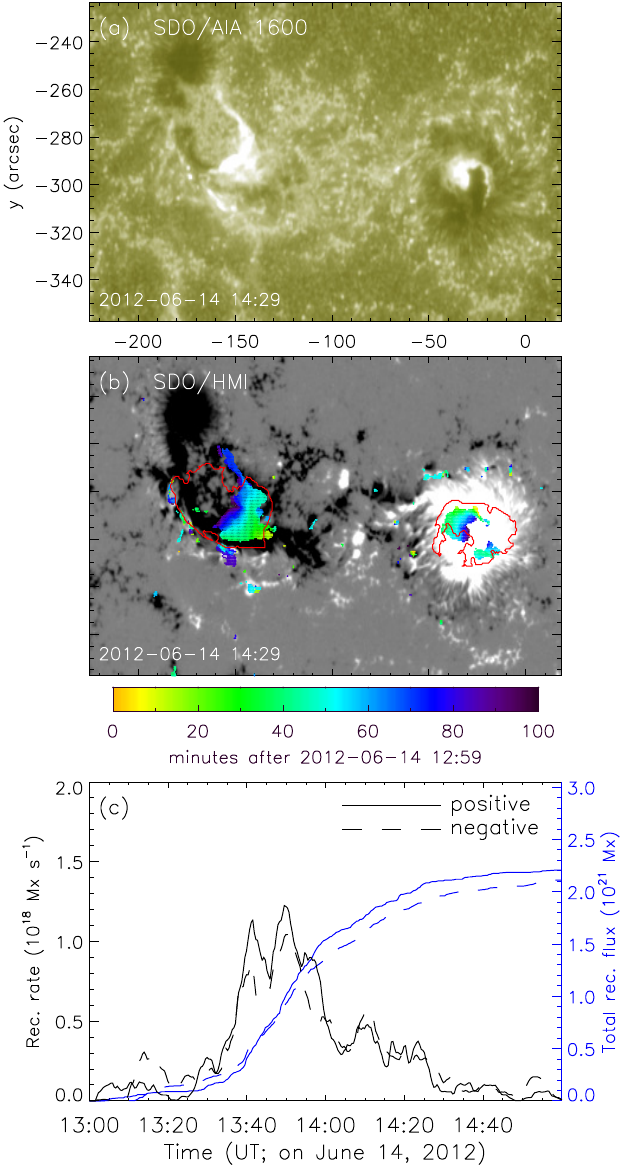}
\caption{The magnetic reconnection properties derived from the associated  flare ribbon morphology: (a) the SDO/AIA 1600 {\AA}   image at the time of the peak of the M1.9 flare, (b) the corresponding HMI magnetogram with the brightened flare ribbon pixels overlaid and color-coded by the elapsed time as indicated by the colorbar, and (c) the corresponding measured accumulative reconnection flux and the reconnection rate for the positive and negative polarities, respectively. In panel (b), the red contours mark the boundaries of the positive and negative polarity footpoint regions of the pre-existing MFR identified by \citet{2019ApJ...871...25W}.  }\label{fig:recflux}
\end{figure}

Table~\ref{tbl:results} summarizes the selected physical properties, mainly concerning the magnetic field topology and flux contents for both the MC and the source region, including the pre-existing MFR as reported by \citet{2019ApJ...871...25W}. The reconnection flux is derived from the temporal and spatial evolution of flare ribbons from our own analysis following the  standard approach of \citet{Qiu2004,Qiu2010}. Such analysis results are presented in Figure~\ref{fig:recflux}. Similar results for the reconnection flux were also provided by \citet{2019ApJ...871...25W}. Depending on how the threshold conditions for the brightening ribbon pixels are chosen, and other factors, the results may differ slightly. Especially considering the time duration during which the brightening pixels are counted and included, the final accumulative magnetic flux enclosed by the areas swept by the brightened ribbon pixels may differ among separate studies. Generally speaking, the longer the duration, the larger the reconnection flux (taken as equivalent to the ribbon flux) becomes. We finish the measurement of the reconnection flux at a time earlier than \citet{2019ApJ...871...25W} when the ribbons started to spread outside of the boundaries of the pre-existing MFR's footpoint regions (the red contours) in Figure~\ref{fig:recflux}b. The accumulative reconnection flux shown in Figure~\ref{fig:recflux}c has gone through a rapid increase phase and is changing more gradually at later times, coinciding with the change in the reconnection rate.  

The results from \citet{2019ApJ...871...25W} indicate that the pre-existing MFR contains significant amount of flux with both footpoints rooted in regions with strong but opposite polarity magnetic field. Considering that for the time period chosen in Figure~\ref{fig:recflux}, the flare ribbons swept through both footpoint regions while largely confined within the boundaries of the pre-existing MFR footpoints as marked. The reconnection flux has to correspond to the reconnected flux between the field lines belonging to the pre-existing MFR. This is consistent with the reconnection sequence identified as ``rope-rope" to ``rope-flare loop" type (or ``rr-rf" in short) by \citet{2019Aulanier}. Simply put, such a scenario may be traced back to the earlier schematics by \citet{1995GeoRL..22..869G}  where the reconnection between two legs of adjacent loops, each leg belonging to a different loop, results in one unit of axial flux being removed due to the disconnection of one pair of footpoints from the loop structure (or the flux rope) above to form a closed flare loop below.  
Therefore a reduction in the axial (toroidal) magnetic flux of the MFR by the amount of the reconnection flux should result. 

This leads to an expected remaining axial magnetic flux of the erupted flux rope in the range of 9 - 20$\times10^{20}$ Mx with about $\gtrsim$20\% uncertainty for this event. The erupted flux rope was later detected by the Wind spacecraft at 1 au with in situ measurements and the FS modeling results yield an axial flux, 8.0
- 14$\times10^{20}$ Mx, which overlaps with the range of anticipated value. In addition, such reconnection may contribute to the apparent knotted feature in Figure~\ref{fig:3D}, which is not present in either 1D or 2D flux rope models. 
The twist number estimates, on average, do not change significantly, although the total twist number estimate for the FS model is still subject to a large uncertainty in field-line length. The magnetic helicity contents are of the same order of magnitudes whereas the value for the MC is smaller due to the reduced axial magnetic flux. 
Moreover, for this event, the connectivity of field lines in the MC flux rope back to the positive polarity footpoint region on the Sun is better maintained, as implied by a larger amount of remaining magnetic flux in the positive polarity footpoint region.
It seems to be consistent with the ePAD data where the relative enhancement of the streaming suprathermal electrons is more pronounced at 0$\deg$ PA  within the MC interval in Figure~\ref{fig:dataplot}, the second panel. This signature, corresponding to unidirectional streaming electrons \citep[see, e.g.,][]{1995GeoRL..22..869G}, generally indicates stronger connectivity of only one end of a field line to the positive polarity footpoint region.

\begin{figure}
\centering
\includegraphics[width=8.3cm]{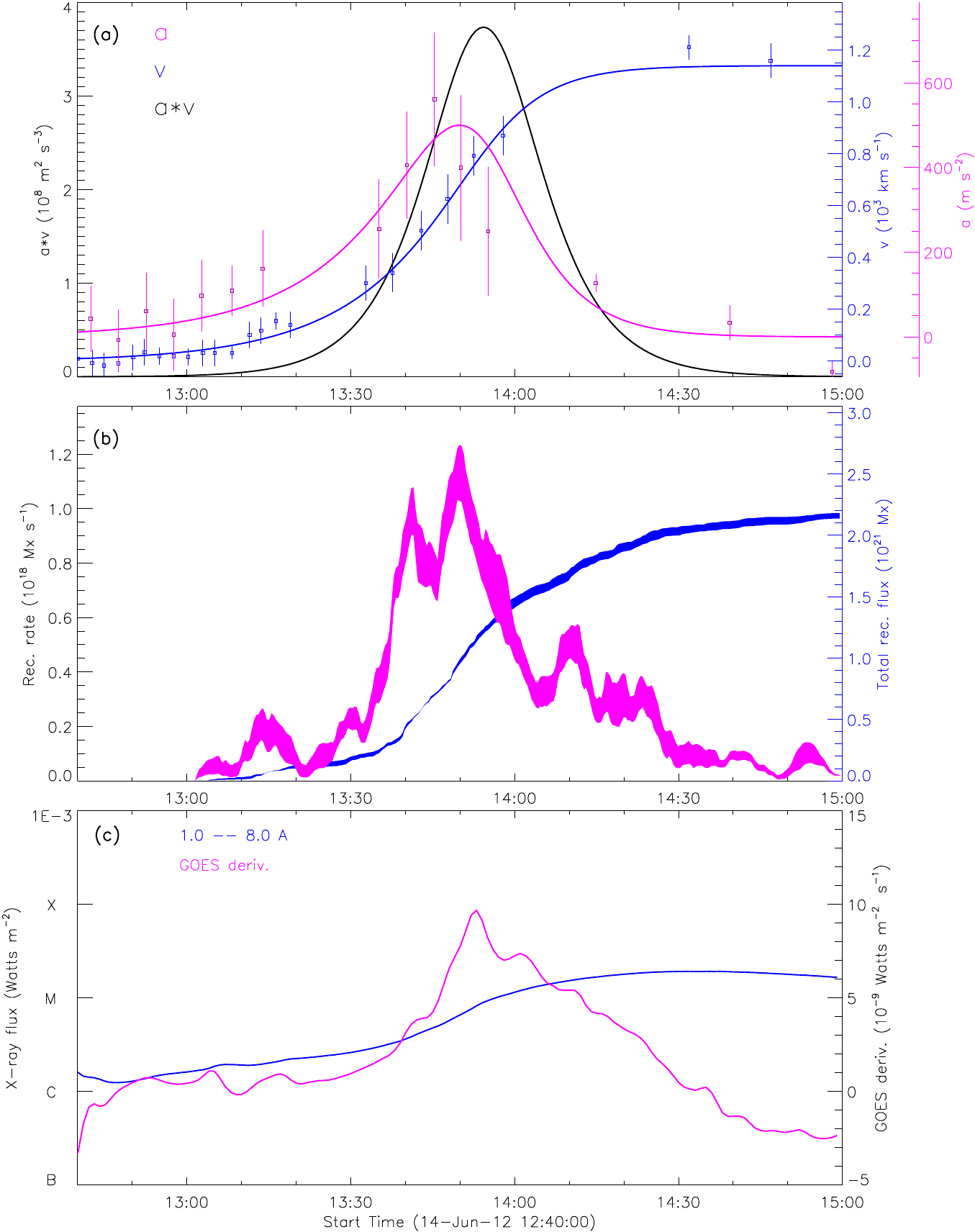}
\caption{Temporal variations of (a) the CME speed $v$, acceleration $a$, and the product $av$, (b) the corresponding reconnection flux and reconnection rate derived from flare ribbons (line thickness represents the range of uncertainty estimates), and (c) the corresponding GOES soft X-ray flux and its derivative. }\label{fig:CMEkin}
\end{figure}

Figure~\ref{fig:CMEkin}a shows the typical set of time-varying  profiles of the speed $v$, and acceleration $a$ of the associated CME \citep{zhu2020}, and the product $av$ as a possible proxy to the rate of change of kinetic energy (assuming little change in CME mass). Figure~\ref{fig:CMEkin}b and c show the corresponding measurements of the accumulative reconnection flux, the rate of change of the reconnection flux, the soft-X ray flux and its temporal derivative, respectively. It shows a general pattern of coincidence in the peaks of the acceleration, the reconnection rate, and the rate of change of the soft-X ray flux, which has been demonstrated for many flare/CME events \citep[e.g.,][]{zhu2020}. The peak in the product $av$ seems to slightly lag behind the other peaks \citep[see, e.g.,][]{Karpen_2012}. A latest numerical simulation study    by \citet{2021jiang2} showed better coincidence between the peak of the rate of  change in kinetic energy and that in the reconnection rate. The overall temporal profiles generally coincide with each other among the corresponding changes in the reconnection rate, the CME acceleration, and the time-derivative of the soft-X ray flux.

In summary, based on these quantitative analysis results, we offer 
the main interpretation such that the reconnection  associated with the M1.9 flare on 2012 June 14  reduces the axial flux of the pre-existing MFR upon eruption by around 50\%, while the MFR-CME kinematic behavior is largely unaffected.


\section{Conclusions and Discussion} \label{sec:discussion}
In conclusion, we have carried out additional quantitative analysis of the flare-CME-ICME/MC event during 2012 June 14-17, especially by performing an analysis of the MC configuration through a newly developed quasi-3D FS  fitting approach. The MC observed by the Wind spacecraft possesses an unusually strong magnetic field magnitude (with the maximum reaching 40 nT), which was also crossed by the VEX spacecraft at $r_h\approx 0.7$ au and $6^\circ$ west of the Sun-Earth line. The optimal fitting of the FS model to the Wind in situ magnetic field measurements yields a minimum reduced $\chi^2\approx 1.44$. The field-line configuration shows an indication of writhe or a knot of the main flux bundle winding in the direction $\hat\mathbf{z}=$(56$^\circ$, 150$^\circ$) in terms of the polar and azimuthal angles in the GSE coordinates. The VEX spacecraft crossed the main flux bundle to the west near the ecliptic from the Earth's point of view toward the Sun. The spatial variations from the Wind spacecraft path to that of VEX are significant and intrinsically 3D based on the FS model output. A comparison of the magnetic field components between the FS model output along the VEX spacecraft path and the actual time-series data yields a correlation coefficient $cc=0.86$.

Based on the uncertainty estimates for the fitting parameters of the FS model, the axial magnetic flux (all in the unit of $10^{20}$ Mx hereafter)  of the MC flux rope is estimated to be 8.0 - 14. A pre-existing MFR was identified in \citet{2019ApJ...871...25W} with two footpoint regions rooted in strong magnetic polarity regions and the axial flux amounting to 42 and 30 for the positive and negative polarities, respectively.  The subsequent M1.9 flare exhibited brightened ribbons mostly confined within the boundaries of the identified pre-existing MFR footpoint regions. The corresponding accumulative reconnection flux reaches 22$\pm4.7$ and 21$\pm5.3$ for the positive and negative polarities, respectively. This implies a reduction of the axial magnetic flux of the pre-existing MFR by the total amount of the reconnection flux. Therefore the erupted CME flux rope following the flare would contain the amount of axial flux in the range 9 - 20 with uncertainties, which agrees with the range of the estimated axial flux of the MC flux rope. This result supports the scenario of the reconnection sequence between the field lines of the pre-existing MFR near the two legs, as envisaged by \citet{2019Aulanier}. In addition, the kinematics of the accompanying CME as analyzed through coronagraphic measurements show little distinction from the general pattern in terms of the coincidence among the peaks of the CME acceleration, the reconnection rate, and the rate of change of the soft X-ray flux. In other words, such a presumably unusual process of the removal of the axial magnetic flux from the erupting MFR through flare reconnection appears to have little effect on  the CME kinematics. The main characteristics remain similar to the other more typical process during which the reconnection flux is injected into the erupting MFR forming the CME \citep[see, e.g.,][]{zhu2020}.
Lastly, the ePAD signatures with relatively greater enhancement  at 0$^\circ$ PA during the MC interval seem to be consistent with the aforementioned interpretation as well. 

The complexity or evolution of a flux rope topology is probably not limited to the process of the leg-leg type reconnection only. Additional ribbon brightenings after the main sequence may correspond to other types of reconnection not resulting in the deduction of axial flux. 
For example, the latest numerical simulation by \citet{2021jiang2} demonstrated the buildup of an MFR from magnetic reconnection with the axial flux of the MFR increasing initially in sync with the reconnection flux, then the axial flux started to decrease while the reconnection flux continued to grow. 
Apparently, not all reconnection flux over an extended period of time corresponds to the reduction in the axial flux of the MFR. For example, one recent observational analysis by \citet{Xing_2020} on the evolution of the axial flux in identified MFR footpoints showed modest decrease by about 10-20\%.
The reconnection in later stages may instead correspond to an interchange type of reconnection with either open or closed neighboring flux systems, resulting in the disconnection or ``drifting" of the MFR footpoints \citep{2019Aulanier,2019ApJ...883...96Z} without altering the axial flux, especially when quantified from in situ measurements. 

We  wish to point out the 3D features of the FS model. 
Based on equation~(\ref{eq:Tw}), the constant parameter $\alpha$ has the meaning of number of twist per unit length (subject to a constant proportional factor). So the total twist  for a single field line is proportional to the product of $\alpha$ and the field-line length $L$ from one end to the other. This will be a finite and unique number for any field line with a finite length. For a straight field line, the twist number becomes arbitrary. 
As we pointed out, there are no straight field lines in the configuration presented here. This imposes a challenge for defining a ``center" of the flux rope to be along a straight field line as we usually do for a 2D configuration \citep{Hu2017GSreview,Hu2021a}. Moreover a distinction between the field-line twist per unit length and a total twist number has to be made, and it is generally  more appropriate and accurate for the latter to be compared with the corresponding source region counterpart, e.g., in Table~\ref{tbl:results}. However such a comparison is not feasible because the current flux rope models based on in situ measurements lack the capability to determine the field-line lengths $L$. Therefore the proper evaluation of the total field-line twist based on in situ MC modeling remains challenging. In addition, a 3D field-line configuration also leads to difficulty in defining the poloidal flux of a flux rope, as we discussed in \citet{2022arXiv220103149H}, simply because of the difficulty in defining a central axis.

We are also aware that \citet{2018ApJ...855L..16J} found a pre-existing MFR prior to the eruption of the M1.9 flare through a non-linear force-free field extrapolation. They succeeded in generating an MFR structure in their solution by using an HMI vector magnetogram patch  around 12:24 UT on 2012 June 14. The amount of axial magnetic flux contained in the MFR is 4$\times10^{20}$ Mx, based on their extrapolation result. The total average twist number was estimated to be between 1.35 and 1.88. Based on Table~\ref{tbl:results}, if one assumes that the accumulative reconnection flux were all injected into the erupting CME flux rope, the total axial flux would amount to $\sim20\times10^{20}$ Mx or more, considering uncertainties. This process may comply with the other scenario as discussed before, although this amount seems to be  larger than the upper limit of the range of axial flux contained in the corresponding MC flux rope. It is possible that not all reconnection flux may add to the (axial) flux content of the MC flux rope. However to discern all possible scenarios would require detailed analysis of reconnection sequence combining flare ribbon morphology and magnetic field topology/connectivity, as attempted by \citet{Qiu2009} for one case study. This will be pursued in future studies. 

\acknowledgments The authors (JQ, CZ, and QH)  acknowledge NASA grant 80NSSC18K0622 for partial support. QH acknowledges NASA grants 80NSSC21K1671, 80NSSC21K0003,  
80NSSC19K0276, and  80NSSC17K0016 for additional support. WH and QH acknowledge NSF grants
AGS-1650854, AGS-1954503, AGS-2050340  and the NSO DKIST Ambassador program for support. LKJ thanks the support of NASA LWS and HSR programs. AP would like to acknowledge the support by the Research Council of Norway through its Centres of Excellence scheme, project number 262622, as well as through the Synergy Grant number 810218 459 (ERC-2018-SyG) of the European Research Council.   The Wind spacecraft  data are accessed via the NASA CDAWeb (\url{https://cdaweb.gsfc.nasa.gov/}). The VEX MAG data are publicly available at the Planetary Plasma Interactions (PPI) node of Planetary Data System (\url{https://pds-ppi.igpp.ucla.edu/index.jsp}). We thank Dr. Wensi Wang for her assistance in providing some of her analysis results.

 \bibliography{ref_master3}

\end{document}